      [1999/12/01 v1.4c Il Nuovo Cimento]
\documentclass{cimento}

\usepackage{graphicx}  
\title{Supernovae Shedding Light on Gamma-Ray Bursts}
\author{M. Della Valle\from{ins:x}}

\instlist{\inst{ins:x} INAF-Arcetri Astrophysical Observatory, 
Largo E. Fermi 5, Firenze (Italy)}

\PACSes{\PACSit{97.60}{Supernovae}}
\PACSes{\PACSit{98.70}{Gamma-Ray sources; Gamma-Ray Bursts}}

\newcommand{\lsim}{{\, \lower2truept\hbox{
${< \atop\hbox{\raise4truept\hbox{$\sim$}}}$}\,}}
\newcommand{\gsim}{{\, \lower2truept\hbox{
${> \atop\hbox{\raise4truept\hbox{$\sim$}}}$}\,}}

\begin{document}

\maketitle

\begin{abstract}
We review the observational status of the Supernova (SN)/Gamma-Ray
Burst (GRB) connection. In section 2 we provide a short summary of the
observational properties of core-collapse SNe. In sections 3-6 we
review the circumstantial evidences and the direct observations that
support the existence of a deep connection between the death of
massive stars and GRBs. Present data suggest that SNe associated with
GRBs form a heterogeneous class of objects including both bright and
faint Hypernovae and perhaps also `standard' Ib/c events. In section
7, we provide an empirical estimate of the rate of Hypernovae, for a
``MilkyWay-like'' galaxy, of about $\sim 2.6\times 10^{-4}$
yr$^{-1}$ that may imply the ratio GRB/Hypernovae to be in the range
$\sim 0.03-0.7$. In the same framework we find the ratio GRB/SNe-Ibc
to be $\sim 0.008\div 0.05$. In section 8 we discuss the possible
existence of a lag between the SN explosion and the associated
gamma-ray event. In the few SN/GRB associations so far discovered the SN
explosions and GRB events appear to go off simultaneously. In section
9 we present the conclusions and highlight the open problems that
Swift hopefully will allow us to solve.
\end{abstract}

\section{Introduction}

Gamma Ray Bursts are sudden and powerful flashes of gamma-ray
radiation that occur randomly in the sky at the rate of about one per
day (as observed by the BATSE instrument). The distribution of the
durations at MeV energies ranges from $T\simeq 10^{-3}$~s to about
$10^3$~s and is clearly bimodal (Kouveliotou et al. 1993), with
``long'' bursts characterized by $T>2$~s. In the original discovery
paper, Klebesadel, Strong \& Olson (1973) pointed out the lack of
evidence for a connection between GRBs and Supernovae (SNe), as
proposed by Colgate (1968), but they concluded that ``{\sl
\ldots the lack of correlation between gamma-ray bursts and reported
supernovae does not conclusively argue against such an
association\ldots}''. This point remained a mystery for almost three
decades and only at the end of the 1990s the discovery of GRB
afterglows (Costa et al. 1997, van Paradijs et al. 1997, Frail et. al
1997) at cosmological distances (Metzger et al. 1997) and the
discovery of SN 1998bw in the error-box of GRB 980425 (Galama et
al. 1998) have started shedding light upon the nature of GRB
progenitors.

\section{Core-Collapse Supernovae}

Type-II SNe represent about 70\% of the exploding stars in the
universe (Cappellaro et al. 1999, Mannucci et al. 2005). They have
never been discovered in elliptical galaxies and this led to the idea
that their progenitors should be massive stars $>8-10 M_\odot$
(e.g. Iben \& Renzini 1983) that undergo core-collapse. This fact was
perceived by Baade and Zwicky (1934a, 1934b): {\sl ``a supernova
represents the transition of an ordinary star into a neutron star
consisting mainly of neutrons''}. The gravitational binding energy of
the imploding core ($\sim 10^{53}$ erg) is almost entirely (99\%)
released as neutrinos whereas only $\sim 10^{51}$ erg are converted in
kinetic energy of the ejecta and a very tiny fraction, $\sim 10^{49}$
erg, in luminous energy.

The spectroscopic classification of SNe dates back to Minkowski
(1941): {\sl ``Spectroscopic observations indicate at least two types
of supernovae.  Nine objects form an extremely homogeneous group
provisionally called ``type I''. The remaining five objects are
distinctly different; they are provisionally designed as type II. The
individual differences in this group are large...''}. Particularly
type-II SNe include objects with prominent hydrogen lines while type-I
class is defined by the lack of hydrogen in the spectra. Nowadays we
designated as {\sl type-Ia} those type-I SNe that are characterized by
a strong absorption observed at $\sim 6150$ \AA~ (attributed to the
P-Cyg profile of Si II, $\lambda\lambda 6347, 6371$) and lack of H.
The absence of H, and the fact that these SNe are discovered also in
elliptical galaxies, hint that they arise from a different explosive
phenomenon, such as the thermonuclear disruption of a white dwarf
approaching the Chandrasekhar limit, after accreting material from a
binary companion or coalescing with it (e.g. Livio 2001 and references
therein). The spectroscopic classification of SNe has been
substantially reviewed in the last 20 years (e.g. Filippenko 1997a,
Hamuy 2003) and six distinct classes of core-collapse SNe can be
recognized from their spectra obtained close to maximum light.

{\bf a)} {\sl Normal type II}. SNe with prominent Balmer lines flanked
by P-Cyg profiles ({\sl ``...the spectrum as a whole resembles that of
normal novae in the transition stage...''}, Minkowski 1941). These SNe
are believed to undergo the collapse of the core when the progenitor
star still retains a huge H envelope ($\sim 10-60 M_\odot$, Hamuy
2003).  The most outstanding feature is the H$\alpha$ line flanked by
a P-Cyg profile.  These SNe show a very high degree of individuality
that is reflected in wide range of observed line widths, that
indicates the existence of a significant range in expansion
velocities.

{\bf b)} {\sl Interacting type II}. SNe belonging to this class show
strong H lines in emission without absorptions. Chugai (1997) pointed
out that these SNe undergo a strong interaction with a ``{\sl dense
wind''} generated by the progenitor during repeated episodes of
mass loss prior to exploding (e.g. SN 1994aj Benetti et
al. 1998). Their spectra are dominated by a broad H$_\alpha$ line
(FWHM$\sim 10000$ km/s) sometimes superimposed by a narrow emission
component (FWHM$\sim 200-300$ km/s). In this case the SN is dubbed as
II-n (``{\sl n}arrow'', Schlegel 1990). Sporadically a narrow P-Cyg
profile can be observed as in Benetti et al. (1999) and in this case
the designation II-d indicates ``{\sl d}ouble profile''.
  

SNe belonging to class {\bf a)} and {\bf b)} can be crudely grouped
into two photometric varieties: type II-P ({\sl plateau}) and type
II-L ({\sl linear}) (Barbon, Ciatti \& Rosino 1979, Doggett \& Branch
1985). Shortly after the explosion all SN light curves are powered by
the radioactive decay of $^{56}{\rm Ni} \to ^{56}{\rm Co}$ and later
on by $^{56}{\rm Co} \to ^{56}{\rm Fe}$, therefore the respective
luminosities are mostly determined by the amount of $^{56}$Ni that has
been synthesized during the SN explosion. The magnitudes at maximum of
type-II SNe span a range of about 5 mag (see Patat et al. 1994), which
may imply that the amount of $^{56}{\rm Ni}$ produced in SN explosions
varies by a factor 100 ($\sim 0.002\div 0.2 M_\odot$, e.g. Turatto et
al. 1998).


{\bf c)} {\sl Type-Ib/c}. This class of SNe was first noted by Bertola
(1964) {\sl ``It is well known that one of the most conspicuous
features in the visual spectrum of type I supernovae is a very deep
absorption at $\lambda$ 6150 \AA...Such absorption is missing in the
present SN\footnote{SN 1962L in NGC 1073}''}. The criteria to classify
the members of this subclass of type-I SNe just as 'peculiar' objects
were adopted in the literature for the following 20 years. Only in the
mid-1980s (Panagia 1985, Elias et al. 1985; Uomoto \& Kirshner 1985;
Wheeler \& Levreault 1995) it was realized that sufficient
observational differences did exist to justify having two separate
classes of objects (see Matheson et al. 2001 for a recent
comprehensive study). Type-Ib SNe are characterized by spectra with no
presence of H (Fig. 1 left panel) or very weak lines (Branch et
al. 2002) and strong He I lines at $\lambda\lambda$ 4471, 5876, 6678
and 7065 \AA~ (Porter \& Filippenko 1987). Type-Ic (Wheeler \&
Harkness 1986) are characterized by weak or absent H and Si II
$\lambda\lambda 6347, 6371$ lines, and no prominent (if not totally
absent) He lines. They show Ca II H \& K, NIR Ca II triplet and O I
lines with P-Cyg profiles. Type-Ib/c SNe have been so far observed
only in late type galaxies and their most outstanding spectroscopic
feature is the lack of H in the spectra. Both facts suggest that their
progenitors are massive stars, possibly in binary systems
(e.g. Mirabel 2004, Maund et al. 2004), which undergo the collapse of
their cores after they have lost the respective H or He envelopes, via
strong stellar wind or transfer to a binary companion via Roche
overflow. This scenario is fully consistent with observations at radio
wavelengths that reveal the existence of a strong radio emission due
to the interaction of the ejecta with a dense pre-explosion stellar
wind ($10^{-5/-6}$ M$_\odot$ yr$^{-1}$)/established circumstellar
medium, produced by the progenitor (Weiler et al. 2002).

\begin {figure}[!h]
\centering
\begin{center}
\includegraphics [width=6.6cm, height=7.0cm,angle=0]{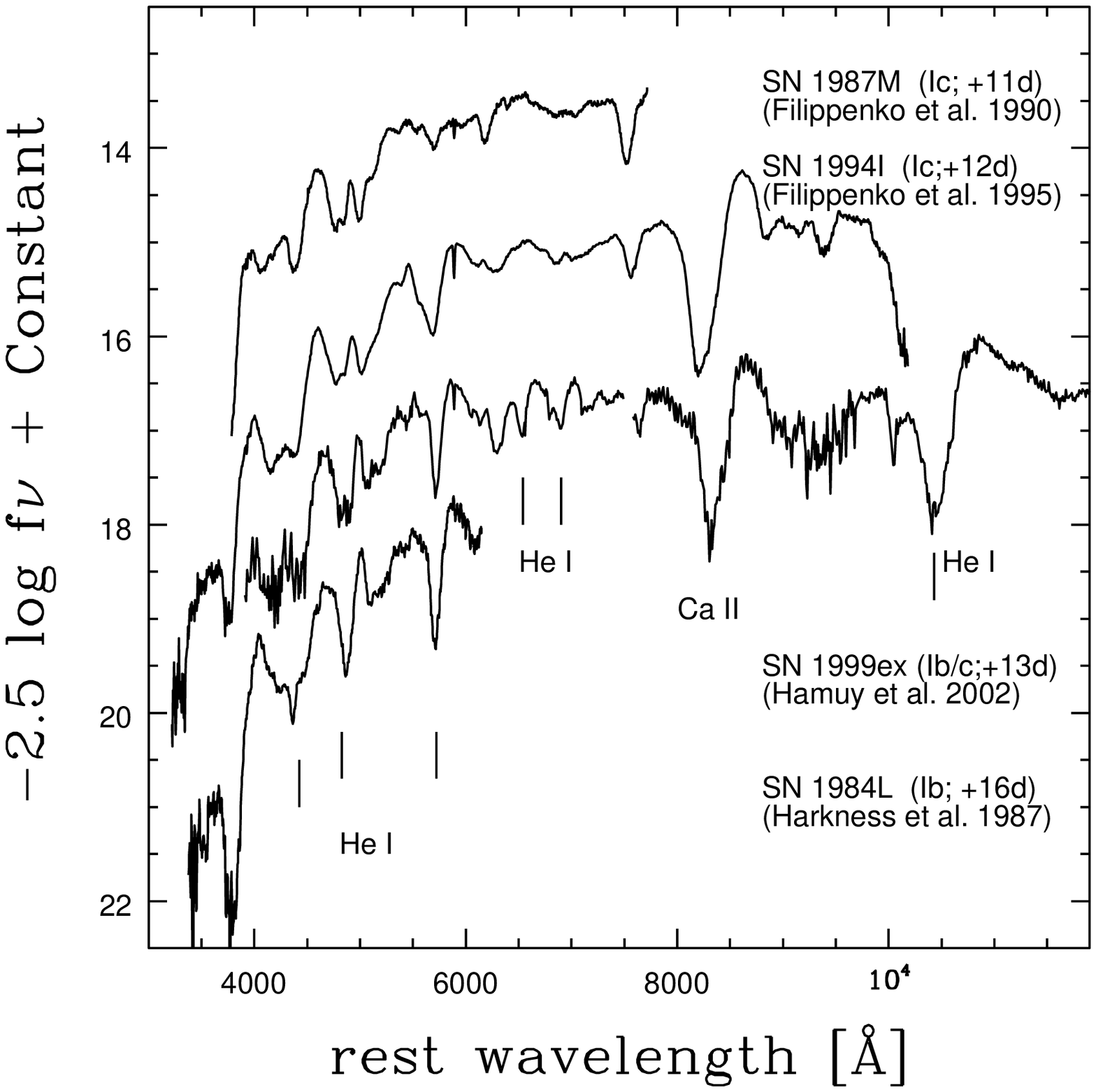}
\includegraphics [width=6.6cm, height=7.0cm,angle=0]{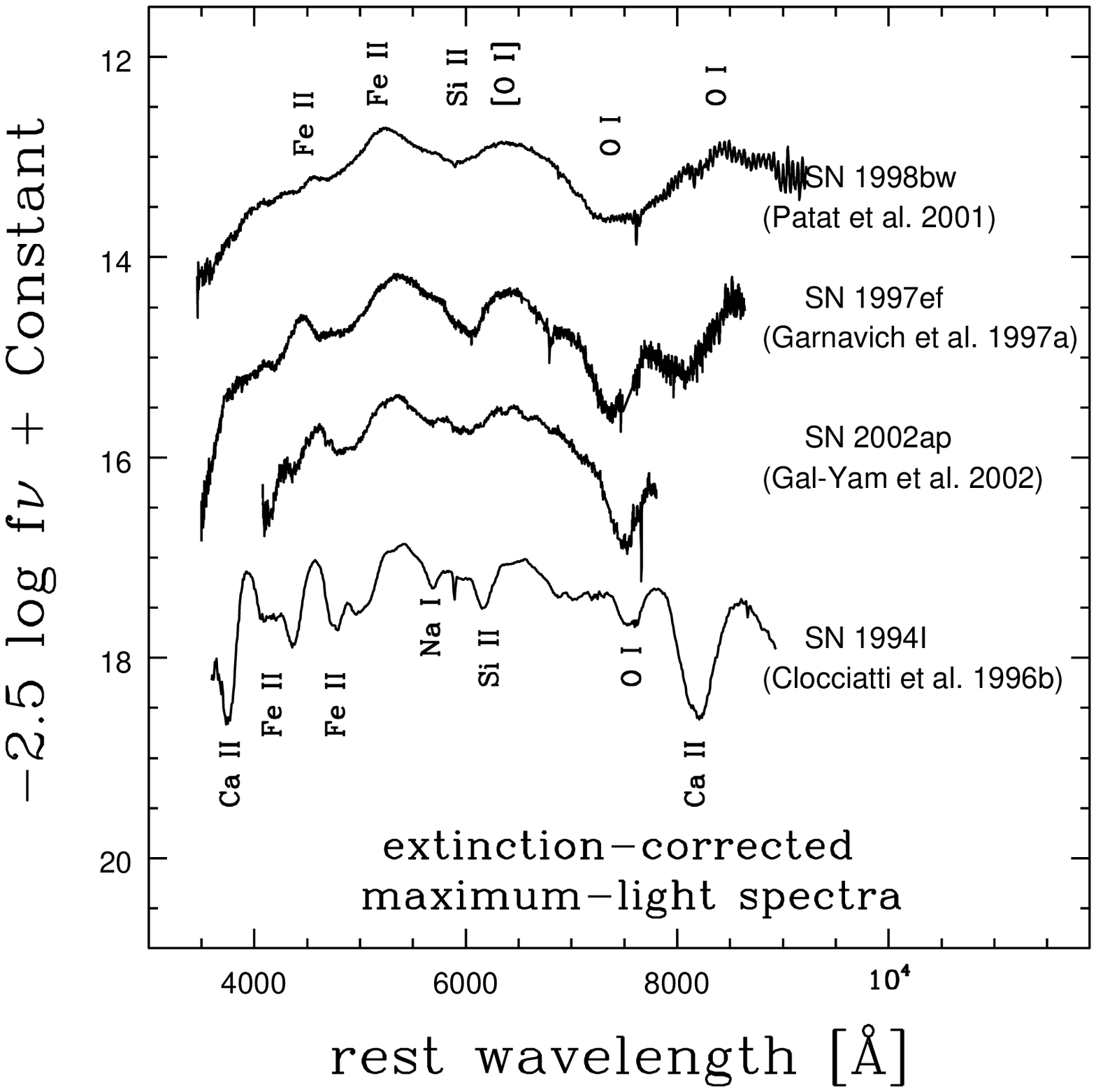}
\caption{{\bf Left panel.} Optical spectra of type-Ib and type-Ic SNe 
obtained a few weeks past maximum.
{\bf Right panel.} Optical spectra at maximum light of Hypernovae, 
compared with the prototypical SN-Ic 1994I. Plots from Hamuy 2003.}
\label{figura}
\end{center}
\end{figure}

{\bf d)} {\sl Type-IIb}. Members of this meagre class of SNe are
objects that evolve from type-II, as inferred by observations of H
lines in the spectra at maximum light, into type Ib (lack of H lines
during the nebular stage, for example). The precursors are thought to
be massive stars that still retain a thin H envelope prior to
exploding.  Prototypical objects of this SN class are SN 1987K
(Filippenko 1988) and SN 1993J (Swartz et al. 1993; Filippenko,
Matheson \& Hot 1993). The discovery of these SNe provides a robust
piece of evidence for the existence of a continuous sequence of SNe
having decreasing envelope mass, i.e. type II--IIb--Ib--Ib/c--Ic.

{\bf e)} {\sl Hypernovae} (Fig. 1, right panel). In 1998 a `weird' SN
(1998bw) was discovered to be spatially and temporally coincident
(chance probability $P\sim 10^{-4/-5}$) with the Gamma Ray Burst
980425 (Galama et al. 1998). The early spectroscopic classification of
this object was not an easy task, indeed. The spectrum at maximum
light was: {\sl ``...not typical of type-Ic supernovae; indeed, the
spectrum does not match any of the known spectral classes, but perhaps
`peculiar type Ic' is the best choice at this time''} (Filippenko
1998). Only observations at late stages (Patat et al. 2001, Stathakis
et al. 2000, Sollerman et al. 2000) have allowed to unambiguously
classify SN 1998bw as type Ib/c. To date (January 2005) 10 more SNe
(see tab. II) have been found to display spectra, at maximum light,
with similar characteristics, i.e. an almost featureless continuum,
characterized by broad undulations, without H lines. The comparison
with the prototypical Ic event, SN 1994I (Filippenko et al. 1995,
Clocchiatti et al. 1996), has confirmed that these `peculiar' objects
are SNe-Ib/c with expansion velocities of the order of 30-40000 km/s
that lead to large Doppler broadening and blending of the emission
lines. The very high expansion velocities of the ejecta suggest that
these SNe are hyper-kinetic (Nomoto et al. 2001), although not
necessarily hyper-luminous, indeed exhibiting a broad range of
luminosities, $M_B\sim -17.0 \div-19.5$, at maximum light (see Fig
2). These objects have been dubbed as {\sl Hypernovae} (note that this
designation was used by Paczy\`nski (1998) with a different meaning).

\begin {figure}[!h]
\centering
\begin{center}
\includegraphics [width=6.6cm, height=7.0cm,angle=0]{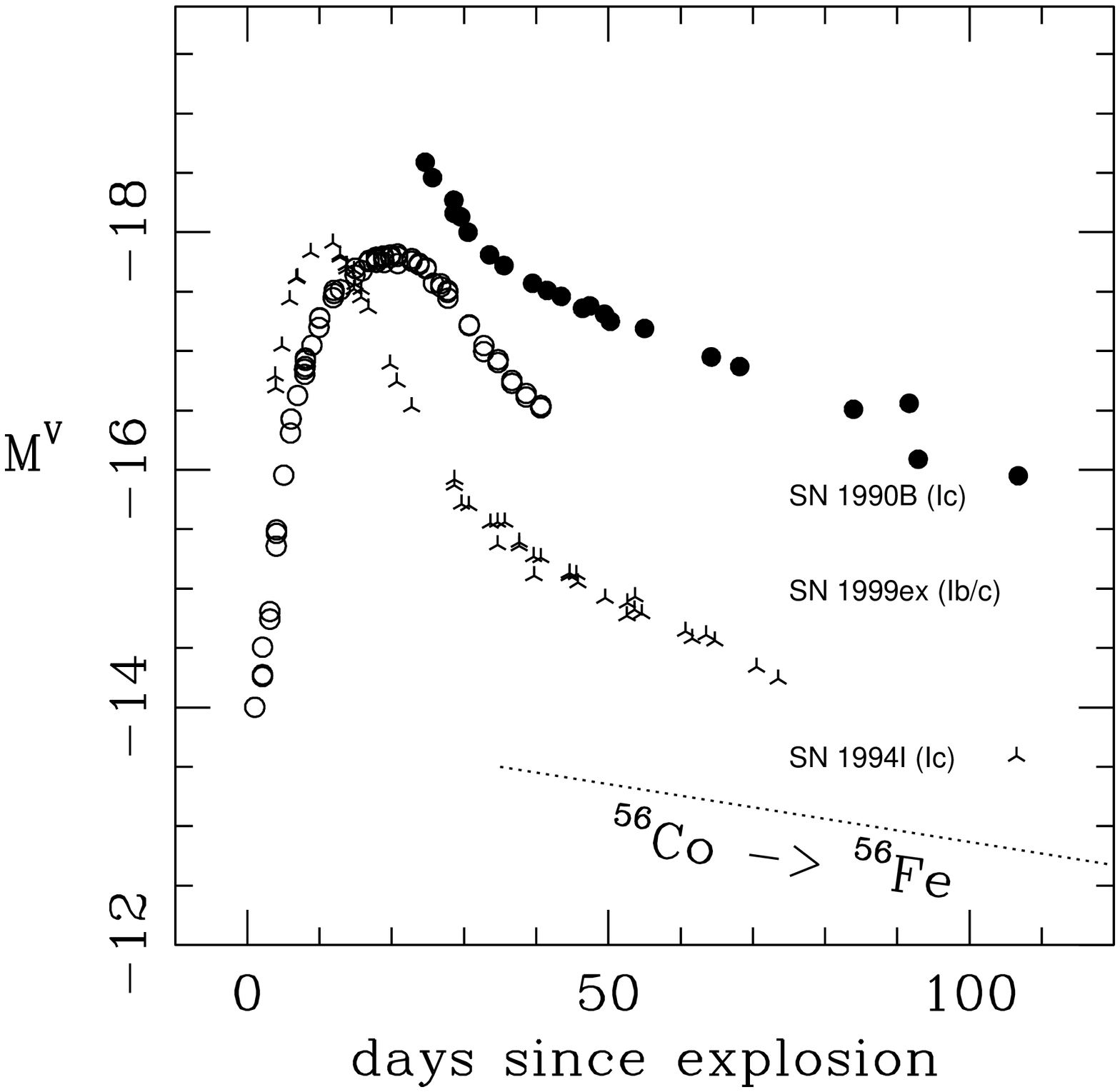}
\includegraphics [width=6.6cm, height=7.0cm,angle=0]{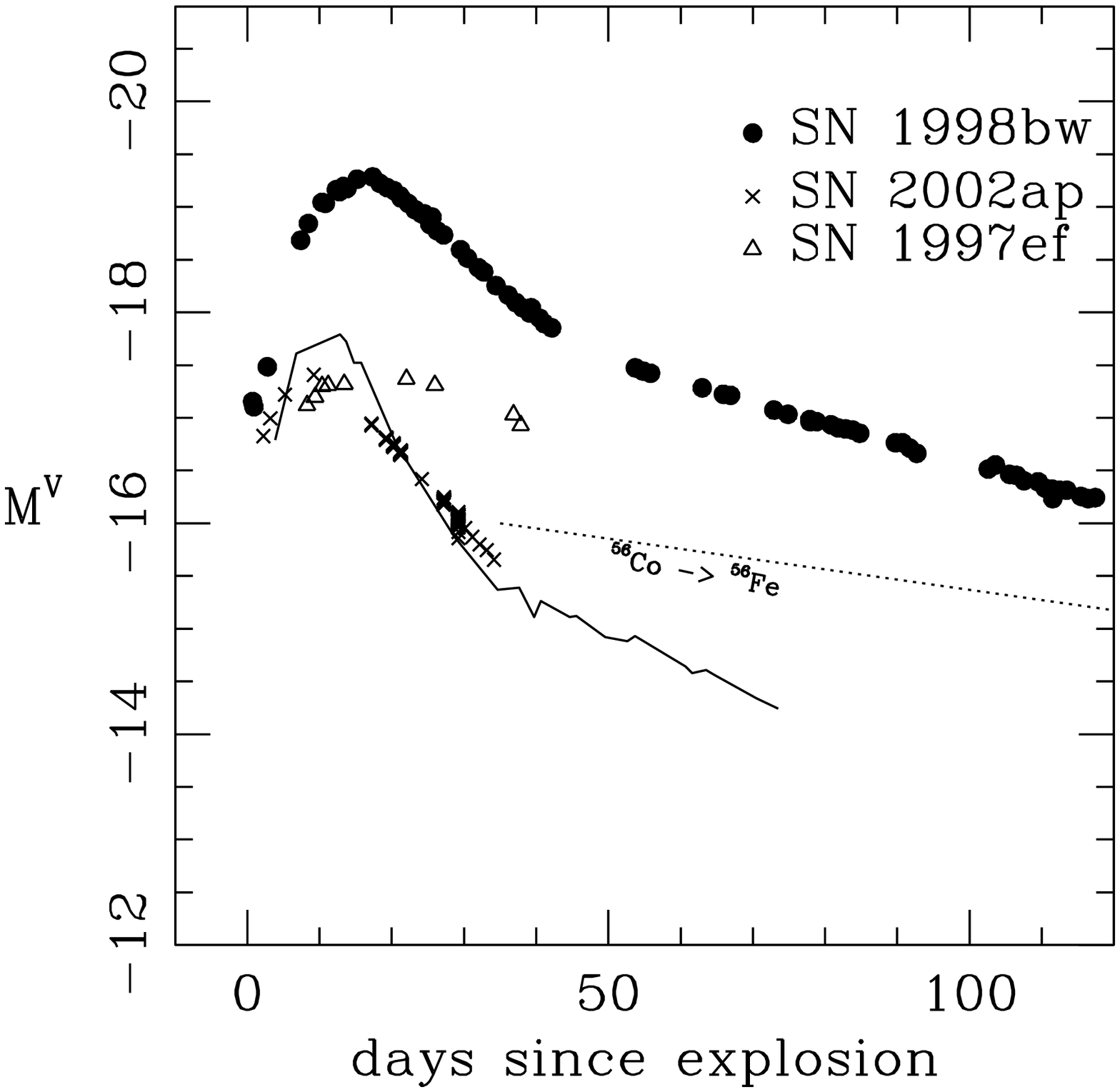}
\caption{{\bf Left panel.} Absolute $V$ light curves of SNe-Ib/c.
Data from Clocchiatti et al. (2001), Richmond et al. (1996) and
Stritzinger et al. (2002). The dotted line represents the expected
trend of the Co-Fe decay. {\bf Right panel.} Absolute $V$ light curves of
Hypernovae, compared to the prototypical SN-Ic 1994I. From the plot is
apparent that Hypernovae are not always overbright with respect to
normal SNe-Ic.  Data from Garnavich et al. (1997a and 1997b), Galama
et al. 1998, McKenzie \& Schaefer (1999), Sollerman et al. (2000),
Patat et al. (2001), Gal-Yam et al. 2002. Plots from Hamuy 2003.}

\label{figura}
\end{center}
\end{figure}

\section{The SN/GRB Connection: Circumstantial Evidences}

Before 2003 the existence of a connection between SNe and long
duration GRBs was supported by several lines of evidence,
even if none of them was really conclusive.
\smallskip

{\bf 1)} SN\,1998bw was the first SN discovered spatially and
temporally coincident with a GRB (GRB\,980425; Galama et
al. 1998). Unexpectedly, SN\,1998bw was discovered not at cosmological
distances, but in the nearby galaxy ESO\,184-G82 at $z = 0.0085$. This
implied that GRB\,980425 was underenegetic by 4 orders of magnitudes
with respect to typical ``cosmological GRBs''. Moreover, the absence
of a conspicuous GRB afterglow contrasted with the associated SN,
which was extremely energetic, had expansion velocities a factor 3-4
higher than those of normal Ib/c SNe and was characterized by a peak
luminosity of $\sim 10^{43}$ erg s$^{-1}$ (for a distance to SN 1998bw
of $\sim 40$ Mpc). This is about 10 times brighter than typical SNe
Ib/Ic (Clocchiatti \& Wheeler 1997), therefore suggesting that a large
amount of $^{56}{\rm Ni}$ must have been synthesized in the SN explosion
(Iwamoto et al. 1998; Woosley, Eastman, \& Schmidt 1999). The
theoretical modeling of the light curve and spectra suggests that SN
1998bw can be well reproduced by an extremely energetic explosion of
an envelope-stripped star, with a C+O core of about $\sim 10 M_\odot$,
which originally was $\sim 40 M_\odot$ on the main sequence (see
Tab. I). This picture is consistent with the radio properties of SN
1998bw, which can be explained as due to the interaction of a mildly
relativistic ($\Gamma \sim$ 1.6) shock with a dense circumstellar
medium (Kulkarni et al. 1998, Weiler et al. 2002) due to a massive
progenitors that has entirely lost its H envelope.

\begin{table}
  \caption{SN 1998bw}
  \label{}
  \begin{narrowtabular}{2cm}{llllll}
     E$_K$($10^{52}$)erg&$^{56}$Ni($M_\odot$)&$M_{core}$($M_\odot$)&$M_{MS}$($M_\odot$)&$M_{left}$($M_\odot$)&Ref.\\
    
    \hline
	2	& 0.7	& 12-15 & 40   & $\sim 2.9$  & Iwamoto et al. 1998\\
	2	& 0.45	& 6-11  & 25-35& $\sim 2$    & Woosley et al. 1999\\
	0.7-5   & 0.4   & 14    & 40   & $\sim 3$    & Nakamura et al. 2001\\     
\hline
  \end{narrowtabular}
\end{table}

H\"oflich, Wheeler \& Wang (1999) presented an alternative picture based
on the hypothesis that all SNe-Ic are the results of aspherical
explosions.  In this case the apparent luminosity of the SN may vary
up to 2 mag, according to different combinations of the geometry of
the explosion and line of sight of the observer. This result can
explain the high luminosity at maximum of SN 1998bw, without calling
for a dramatic overproduction of $^{56}$Ni ($\sim 0.2$ $M_\odot$
$^{56}$Ni) and would allow SN 1998bw to have an explosion energy
($\sim 2\times 10^{51}$ erg) similar to that of `normal' core-collapse
supernovae. Maeda et al. (2002), after analyzing the line profiles in
late time spectra of SN 1998bw, also give some support to the idea
that SN 1998bw was the product of an asymmetric explosion viewed from
near the jet direction (yet characterized by high kinetic energy, of
$\sim 10^{52}$ erg). The idea that Hypernovae and more generally
SNe-Ib/c can be produced by asymmetric explosions is supported by
polarimetry observations of core-collapse SNe (e.g. Wang et al. 2001,
Leonard et al. 2000), which seem to indicate that the degree of
polarization increases along the SN-type sequence: II$\to$ Ib$\to$ Ic
(i.e. with decreasing the envelope mass).
\smallskip

However, the association between two peculiar astrophysical objects
such as GRB 980425 (very faint gamma-ray emission, unusual afterglow
properties) and SN 1998bw (overluminous SN characterized by unusual
spectroscopic features) was believed to be only suggestive, rather than
representative, of the existence of a general SN/GRB connection.
\smallskip

{\bf 2)} The light curves of many afterglows show rebrightenings that
have been interpreted as emerging supernovae outshining the afterglow
several days or weeks after the GRB event (Bloom et al. 1999, Zeh,
Klose \& Hartmann 2004, and references therein). However, since other
explanations such as dust echoes (Esin \& Blandford 2000) or thermal
re-emission of the afterglow light (Waxman \& Draine 2000) could not
be ruled out, only spectroscopic observations during the rebrightening
phase could remove the ambiguity. Indeed spectroscopic features of SNe
are unique, being characterized by FWHM $\sim 100$~\AA~ (see section
4).

{\bf 3)} The detection of star--formation features in the host
galaxies of GRBs (Djorgovski et al. 1998, Fruchter et al. 1999) has
independently corroborated the existence of a link with the death of
massive stars.  For example, Christensen, Hjorth \& Gorosabel (2004)
have found that GRB hosts are galaxies with a fairly high (relative to
the local Universe) star formation of the order of
10~$M_\odot$~yr$^{-1}/L^\star$ (see also Le Floc'h et al. 2003).  Also
the location of GRBs within their host galaxies seems consistent with
the regions that contain massive stars (Bloom, Kulkarni \& Djorgovski
2002a).

{\bf 4)} Some GRB afterglows have shown absorption features at
velocities of a few $\times 10^3$ km/s that has been interpreted as
the result of the interaction with the stellar winds originating from
the massive progenitors (Chevalier \& Li 2000, Mirabal et
al. 2003).

\section{SN 2002lt/GRB~021211}

One of the first opportunities to carry out spectroscopic observations
during a GRB afterglow rebrightening arrived in late 2002.
GRB\,021211 was detected by the \mbox{HETE--2} satellite (Crew et
al. 2003), allowing the localization of its optical afterglow (Fox et
al. 2003) and the measurement of the redshift $z=1.006$ (Vreeswijk et
al. 2002).  Fig.~3 shows the result of the late-time photometric
follow-up, carried out with the ESO VLT--UT4 (Della Valle et
al. 2003), together with observations collected from literature. A
rebrightening is apparent, starting $\sim 15$~days after the burst and
reaching the maximum ($R \sim 24.5$) during the first week of
January. For comparison, the host galaxy has a magnitude $R = 25.22
\pm 0.10$, as measured in late-time images. A spectrum of the
afterglow + host obtained with FORS\,2, 27 days after the GRB, during
the rebrightening phase is shown in the rest frame of the GRB
(red solid line). The spectrum of the bump is characterized by broad
low-amplitude undulations blueward and redward of a broad absorption,
the minimum of which is measured at $\sim 3770$~\AA{} (in the rest
frame of the GRB), whereas its blue wing extends up to $\sim
3650$~\AA. The comparison with the spectra of SN\,1994I, and to some
extent also of SN\,1991bg and SN\,1984L (Fig. 2 in Della Valle et
al. 2003) supports the identification of the broad absorption with a
blend of the Ca\,II $H$ and $K$ absorption lines. The blueshifts
corresponding to the minimum of the absorption and to the edge of the
blue wing imply velocities $v \sim 14\,400$~km/s and $v
\sim 23\,000$~km/s, respectively. The exact epoch when the SN exploded
depends crucially on its rising time to maximum light.  SN\,1999ex, SN
1998bw and SN\,1994I (the best documented examples of type-Ic SNe)
reached their $B$-band maximum $\sim 18$, 16 and 12~days after the
explosion (Hamuy 2003). In Fig.~3 the light curve of SN\,1994I
(dereddened by $A_V = 2$~mag) is added to the afterglow and host
contributions, after applying the appropriate $K$-correction (solid
line). As it can be seen, this model reproduces well the shape of the
observed light curve. A null time delay between the GRB and the SN
explosions is required by our photometric data, even if a delay of a
few days would also be acceptable given the uncertainties in the
measurements. It is interesting to note that SN\,1994I, the spectrum
of which provides the best match to the observations, is a typical
type-Ic event rather than a bright {\sl Hypernova} as the ones
proposed for association with other long duration GRBs (Galama et al. 1998;
Stanek et al. 2003; Hjorth et al. 2003; Malesani et al. 2004).

\begin {figure}[!h]
\centering
\begin{center}
\includegraphics [ width=6.2cm, height=7.2cm, angle=0]{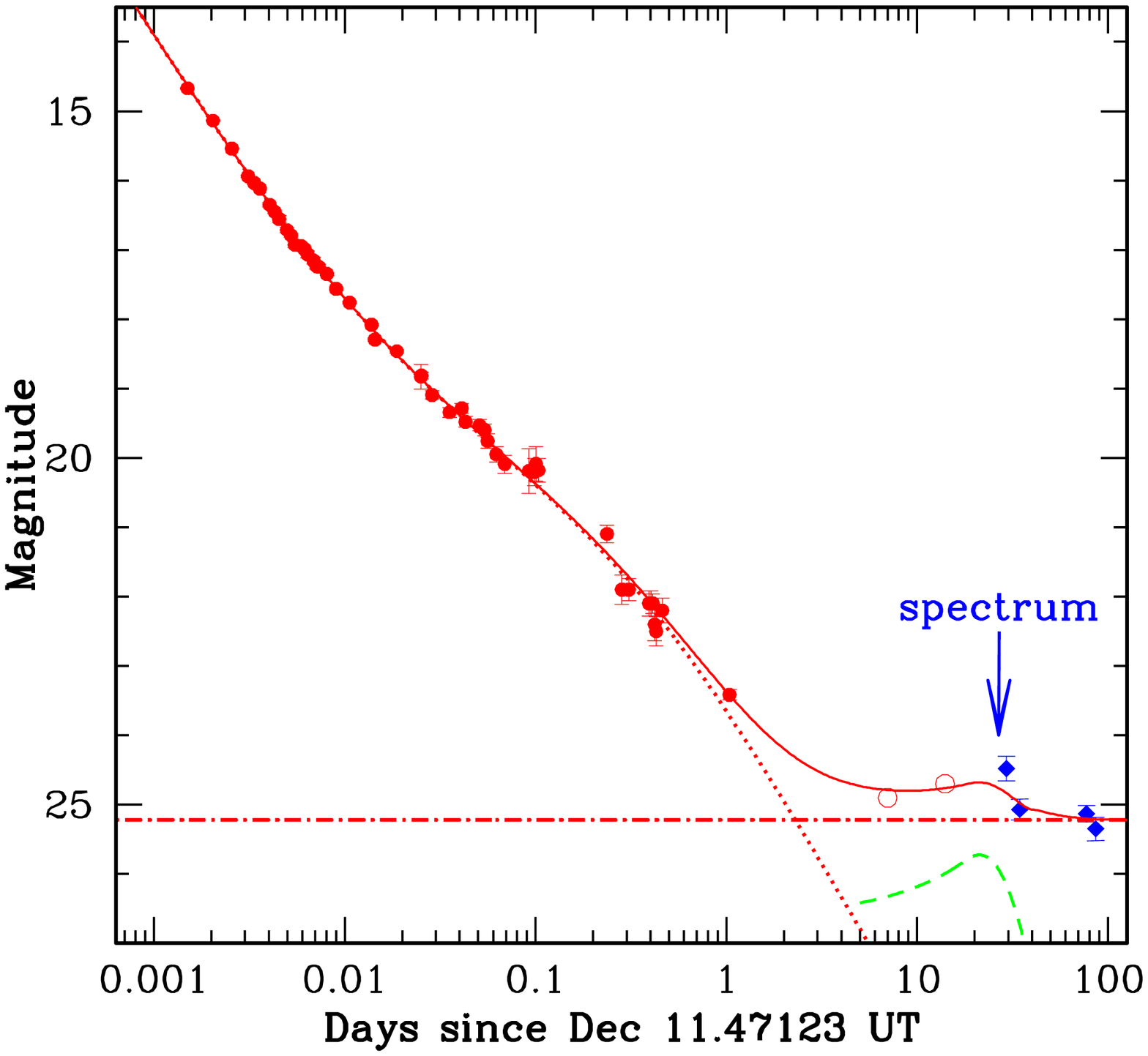}
\includegraphics [width=6.9cm, height=8.2cm,angle=0]{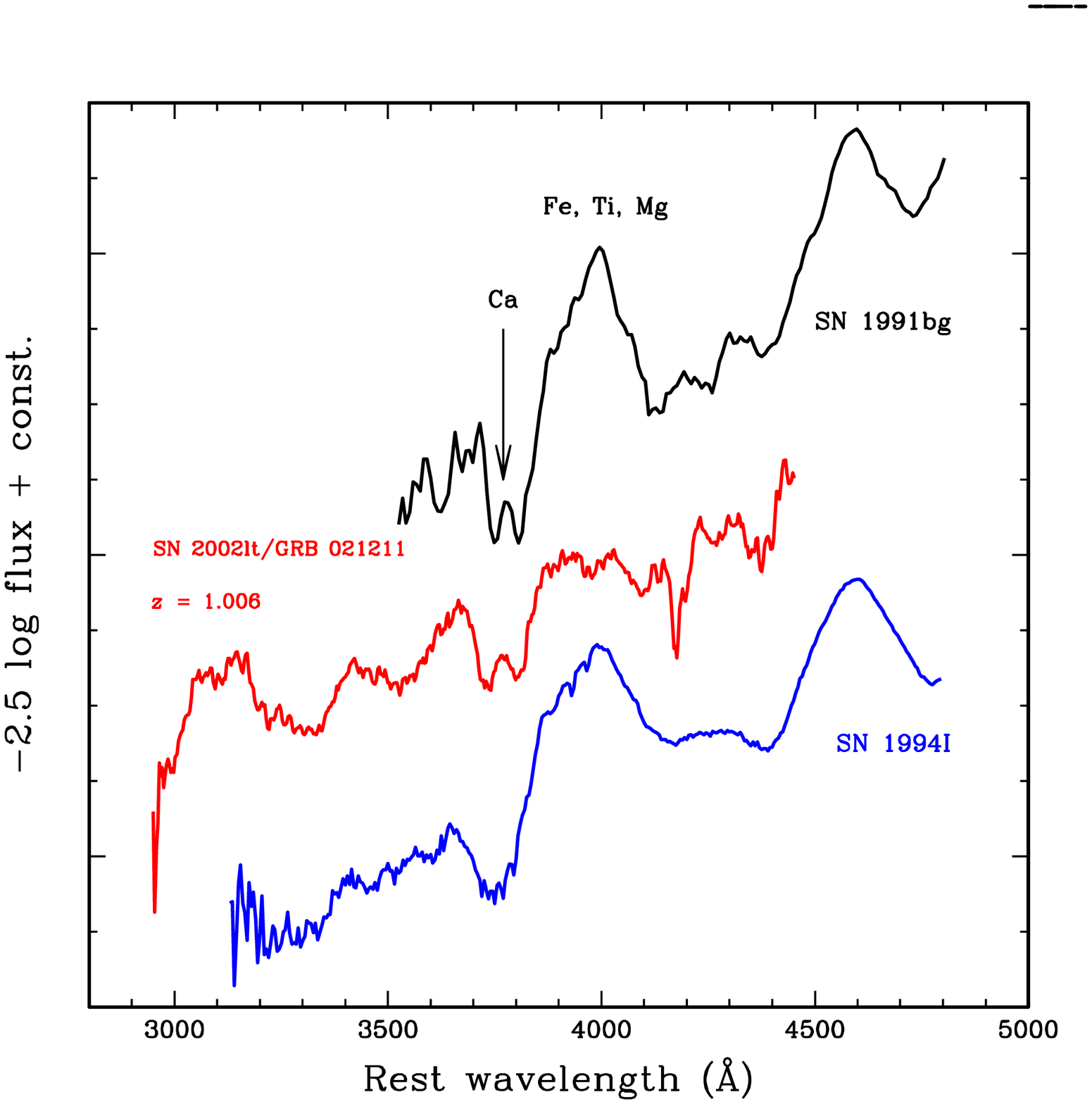}
\caption{{\bf Left panel.} Light curve of the afterglow of GRB 021211. 
Filled circles represent data from published works (Fox et al. 2003;
Li et al. 2003; Pandey et al. 2003), open circles are converted from
HST measurements (Fruchter et al. 2002), while filled diamonds
indicate our data; the arrow shows the epoch of our spectroscopic
measurement. The dotted and dot-dashed lines represent the afterglow
and host contribution respectively. The dashed line shows
the light curve of SN 1994I reported at $z=1.006$ and dereddened with
$A_V=2$ (from Lee et al. 1995). The solid line shows the sum of the
three contributions.  {\bf Right panel.} Spectrum of the afterglow+host
galaxy of GRB 021211 (middle line), taken on Jan 8.27 UT (27 days
after the burst). For comparison, the spectra of SN 1994I (type Ic,
bottom) and SN 1991bg (peculiar type Ia, top) are displayed, both
showing the Ca absorption. Plots from Della Valle et al. 2003, 2004.}
\label{figura}
\end{center}
\end{figure}

\section{The ``Smoking Gun'': GRB\,030329/SN\,2003dh}

The peculiarity of the SN\,1998bw/GRB\,980425 association and the
objective difficulties to collect data for SN\,2002lt (4~h at VLT to
get one single spectrum) prevented us from generalizing the existence
of a SN/GRB connection, although both cases were clearly
suggestive. The breakthrough in the study of the GRB/SN association
arrived with the bright GRB\,030329. This burst, also discovered by
the \mbox{HETE--2} satellite, was found at a redshift $z = 0.1685$
(Greiner et al. 2003), relatively nearby, therefore allowing detailed
photometric and spectroscopic studies. SN features were detected in
the spectra of the afterglow by several groups (Stanek et al. 2003,
Hjorth et al. 2003; see also Kawabata et al. 2003; Matheson et
al. 2003a) and the associated SN (SN\,2003dh) looked strikingly
similar to SN\,1998bw (Fig. 4). The gamma-ray and afterglow properties
of this GRB were not unusual among GRBs, and therefore, the link
between GRBs and SNe was eventually established to be general, so
that, it applies to all ``classical'' and ``long'' cosmological GRBs.

\begin {figure}[!h]
\centering
\begin{center}
\includegraphics [width=6.9cm, height=8.2cm,angle=90]{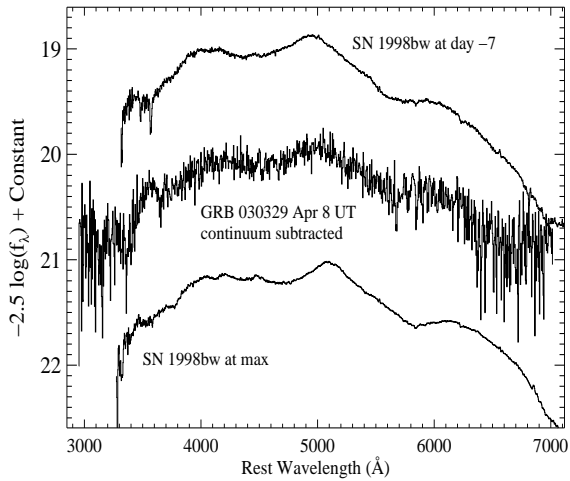}
\caption{Spectrum of 2003 April 8 with the smoothed spectrum of April 4
scaled and subtracted. The residual spectrum shows broad bumps at
approximately 5000 and 4200 \AA~ (rest frame), which is similar to the
spectrum of the peculiar type-Ic SN 1998bw a week before maximum light
(Patat et al. 2001). Plot from Stanek et al. 2003.}
\label{figura}
\end{center}
\end{figure}

The modeling of the early spectra of SN 2003dh (Mazzali et al. 2003)
has shown that SN 2003dh had a high explosion kinetic energy, $\sim
4\times 10^{52}$ erg (if spherical symmetry is assumed). However, the
light curve derived from fitting the spectra suggests that SN 2003dh
was not as bright as SN 1998bw, ejecting only $\sim 0.35$ M$_\odot$ of
$^{56}$Ni. The progenitor was a massive envelope-stripped star of
$\sim 35-40M_\odot$ on the main sequence (Mazzali et al. 2003). The
spectral analysis of the nebular-phase emission lines carried out by
Kosugi et al. (2004) suggests that the explosion of the progenitor of
the GRB 030329 was aspherical, and that the axis is well aligned with
both the GRB relativistic jet and our line of sight.

\section{GRB\,031203/SN\,2003lw: the Older Brother of GRB 980425/SN 1998bw}

GRB\,031203 was a 30s burst detected by the INTEGRAL burst alert
system (Mereghetti et al. 2003) on 2003 Dec 3. At $z = 0.1055$
(Prochaska et al. 2004), it was the second closest burst after
GRB\,980425. The burst energy was extremely low, of the order of
$10^{49}$~erg, well below the ``standard'' reservoir $\sim
10^{51}$~erg of normal GRBs (Frail et al. 2001, Panaitescu
\& Kumar 2001). Only GRB\,980425 and XRF\,020903 were less energetic. In this
case, a very faint NIR afterglow could be discovered, orders of
magnitude dimmer than usual GRB afterglows (Malesani et al. 2004).  A
few days after the GRB, a rebrightening was apparent in all optical
bands (Bersier et al. 2004; Thomsen et al. 2004; Cobb et al. 2004;
Gal-Yam et al. 2004). The rebrightening amounted to $\sim 30\%$ of the
total flux (which is dominated by the host galaxy), and was coincident
with the center of the host galaxy to within $0.1''$ ($\sim
200$~pc). For comparison, in Fig. 5 the $VRI$ light curves of
SN\,1998bw are plotted (solid lines; from Galama et al. 1998), placed
at $z = 0.1055$ and dereddened with $E_{B-V} = 1.1$. Even after
correcting for cosmological time dilation, the light curve of
SN\,2003lw is broader than that of SN\,1998bw, and the latter requires
an additional stretching factor of $\approx 1.1$ to match the $R$ and
$I$ data points. The $R$-band maximum was reached in $\sim 18$
(comoving) days after the GRB. After assuming a light curve shape similar to
SN\,1998bw, which had a rise time of 16~days in the $V$ band, data
suggest an explosion time nearly simultaneous with the GRB. A precise
determination of the absolute magnitude of the SN is made difficult by
the uncertain extinction.  Based on the ratios of the Balmer lines of
the host galaxy, the average combined Galactic and host extinction is
$E_{B-V} \approx 1.1$. Given the good spatial coincidence of the SN
with the center of the host, such value is a good estimate for the SN
extinction. With the assumed reddening, SN\,2003lw appears to be
brighter than SN\,1998bw by 0.5~mag in the $V$, $R$, and $I$
bands. The absolute magnitudes of SN\,2003lw are hence $M_V =
-19.75\pm0.15$, $M_R = -19.9\pm0.08$, and $M_I =
-19.80\pm0.12$. Fig. 5 also shows the spectra of the rebrightening on
2003 Dec ~20 and Dec ~30 (14 and 23 rest-frame days after the GRB),
after subtracting the spectrum taken on Mar ~1 (81 rest-frame days
after the GRB, Tagliaferri et al. 2004). The spectra of SN\,2003lw are
remarkably similar to those of SN\,1998bw obtained at comparable
epochs (shown as dotted lines in Fig.~5, see Malesani et al. 2004 for
details). Both SNe show very broad absorption features, indicating
high expansion velocities. This makes SN\,2003lw another example of
Hypernova. A preliminary analysis of early spectra of 2003lw (Mazzali
et al. 2005, in preparation) indicates that this Hypernova produced a
large amount of Ni, possibly in the range $0.6-0.9 M_\odot$. The
progenitor mass could be as large as 40-50 $M_\odot$ on the main
sequence.
\bigskip

\begin {figure}[!h]
\centering
\begin{center}
\includegraphics [ width=5.9cm, height=7.2cm, angle=0]{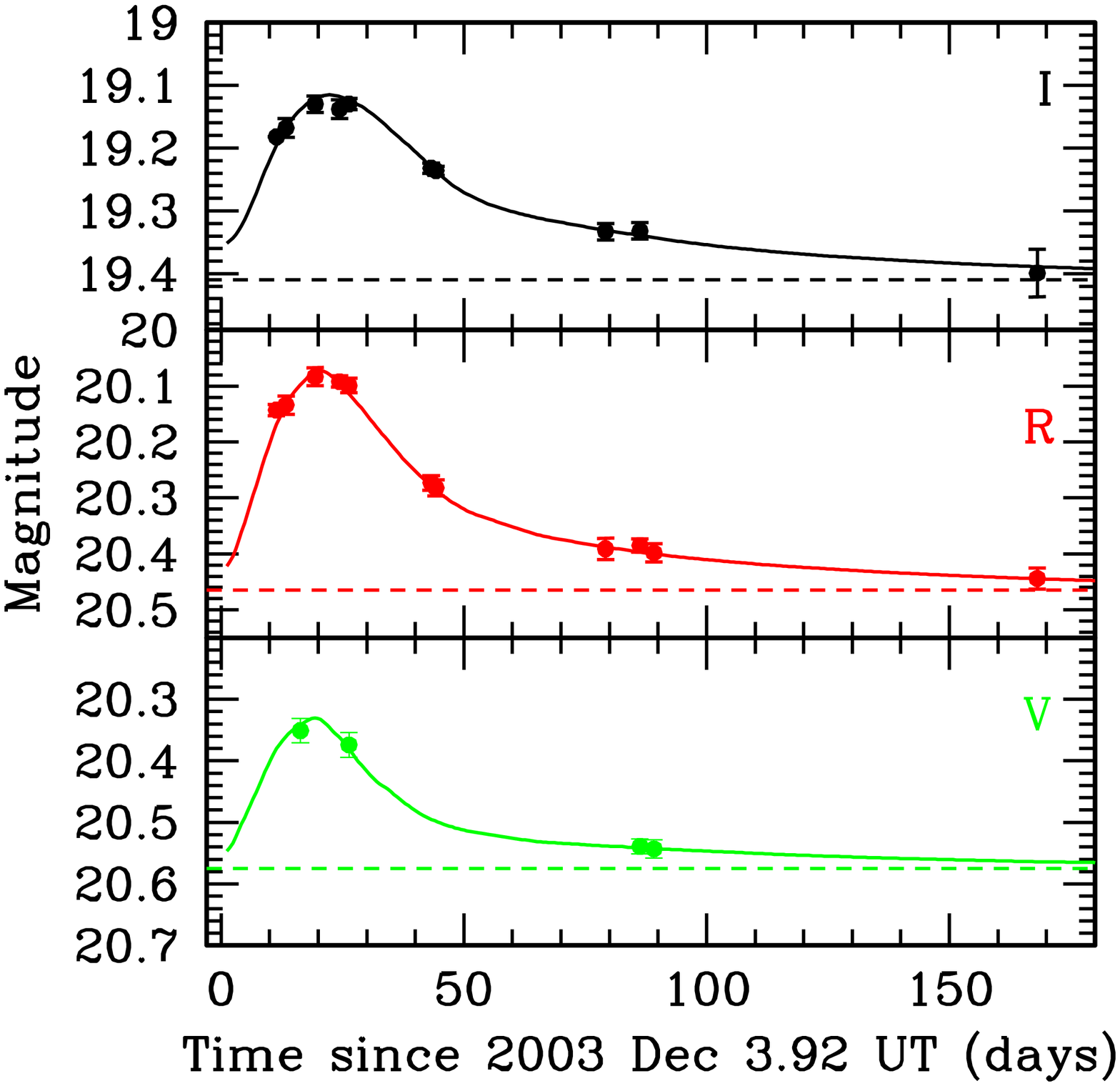}
\includegraphics [width=6.9cm, height=8.2cm,angle=0]{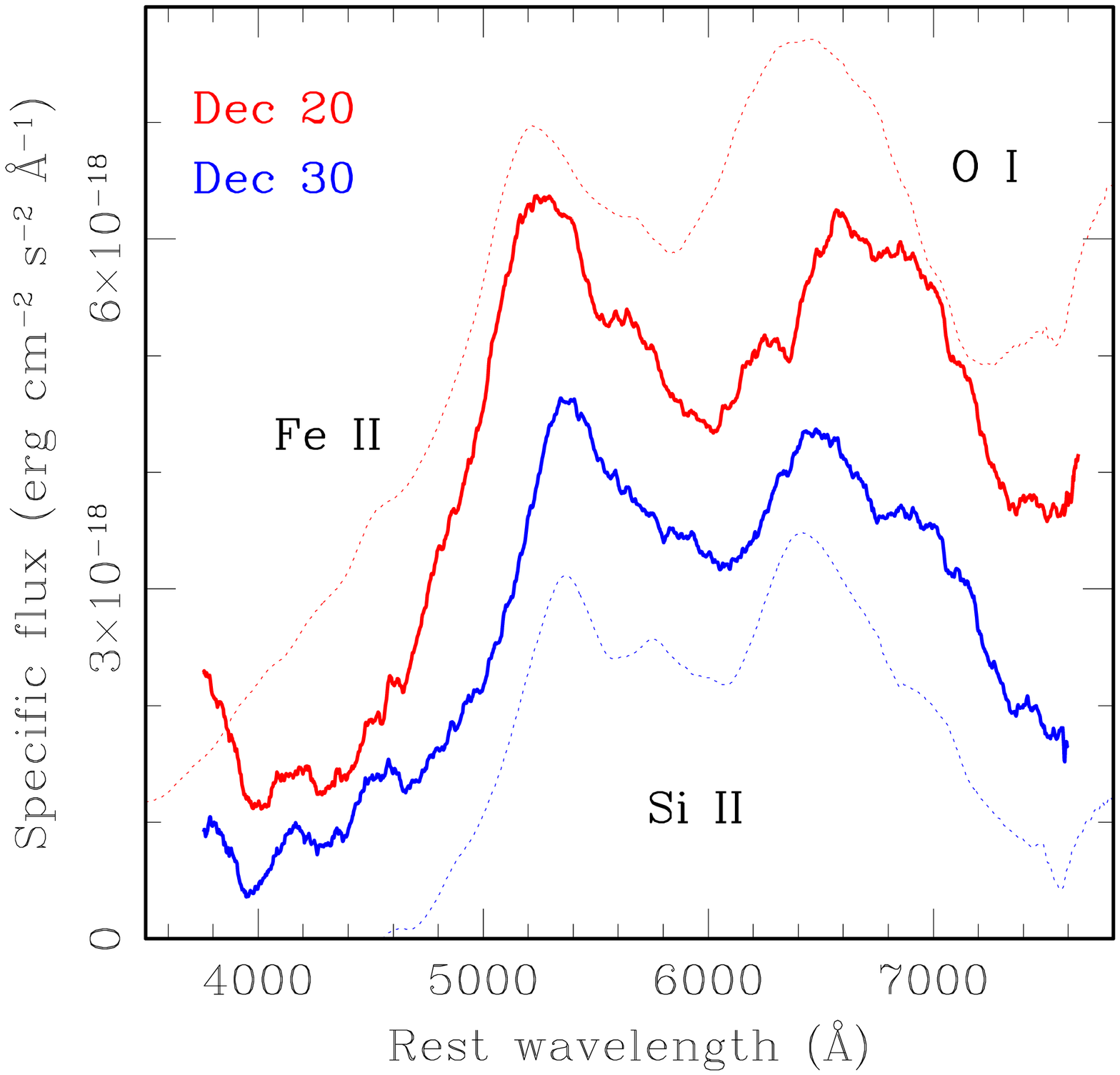}
\caption{{\bf Left panel.} 
Optical and NIR light curves of GRB 031203 (circles). The solid
curves show the evolution of SN 1998bw (Galama et al. 1998; McKenzie
\& Schaefer 1999), rescaled at $z=0.1055$, stretched by a factor 1.1,
extinguished with $E(B-V)=1.1$, and brightened by 0.5 mag. The dashed
lines indicate the host galaxy contribution. The vertical dotted lines
mark the epochs of our spectra. {\bf Right panel.} Spectra of SN
2003lw, taken on 2003 December 20 and 30 (solid lines), smoothed with
a boxcar filter 250\AA~ wide. Dotted lines show the spectra of SN
1998bw (from Patat et al. 2001), taken on 1998 May 9 and 19 (13.5 and
23.5 days after the GRB, or 2 days before and 7 days after the $V$-band
maximum, respectively), extinguished with $E(B-V)=1.1$ and a Galactic
extinction law (Cardelli et al. 1989). The spectra of SN 1998bw were
vertically displaced for presentation purposes. Plots from Malesani et
al. 2004.}
\label{figura}
\end{center}
\end{figure}

\begin{table}
  \caption{Hypernovae}
  \label{}
  \begin{narrowtabular}{2cm}{lrr}
     SN     & $cz$ (km/s)& References \\
    \hline
     1997dq & 958 & Mazzali et al. 2004\\
     1997ef & 3539& Filippenko 1997b\\
     1998bw & 2550& Galama et al. 1998 \\
     1999as & 36000&Hatano et al. 2001\\		
     2002ap & 632 & Mazzali et al. 2002, Foley et al. 2003\\ 
     2002bl & 4757& Filippenko et al. 2002\\
     2003bg & 1320& Filippenko \& Chornack 2003\\
     2003dh & 46000&Stanek et al. 2003, Hjorth et al. 2003\\	
     2003jd & 5635& Filippenko et al.2003; Matheson et al. 2003b\\
     2003lw & 30000&Malesani et al. 2004\\
     2004bu & 5549& Foley et al. 2004 \\
    \hline
  \end{narrowtabular}
\end{table}

\begin {figure}[!h]
\centering
\begin{center}
\includegraphics [width=7.9cm, height=9.4cm,angle=270]{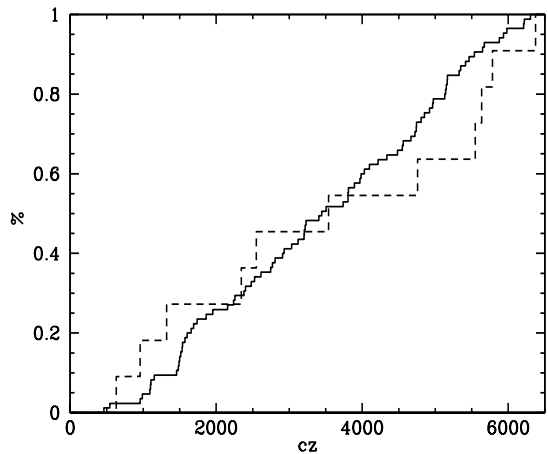}
\caption{Cumulative distribution for SNe Ib/c (solid line) and Hypernovae (dashed line)
discovered in galaxies with radial velocities $cz< 6000 $ km/s.}
\label{figura}
\end{center}
\end{figure}

\section{Rates of SNe Ib/c, Hypernovae and GRBs}
The measurement of the SN rate is based on the control-time
methodology (Zwicky 1938) that implies the systematic monitoring of
galaxies of known distances and the use of appropriate templates for
the light curves of each SN type (see Cappellaro et al. 1993 for bias
and uncertainties connected with this procedure). Unfortunately all
Hypernovae reported in Tab. II have been not discovered during time
`controlled' surveys, and therefore any attempt to derive an absolute
value of the rate of Hypernovae should be taken with great caution.
One possibility is to compute the frequency of occurence of all
SNe-Ib/c and Hypernovae in a limited distance sample of objects. From
the Asiago catalogue (\texttt{http://web.pd.astro.it/supern}) we have
extracted 91 SNe-Ib/c, 8 of which are Hypernovae, with $cz< 6000$
km/s. This velocity threshold is suitable to make the distance
distribution of `normal' Ib/c and Hypernovae statistically
indistinguishable (KS probability=0.42, see Fig. 6). After assuming
that the host galaxies of both `normal' SNe Ib/c and Hypernovae have
been efficiently (or inefficiently) monitored by the same extent, one
can infer that the fraction of Hypernovae is about $7/91\simeq 8\%$
(after excluding SN 1998bw because it was searched in the error-box of
GRB 980425) of the total number of SNe Ib/c. Since Hypernovae can be
brighter than normal SNe-Ib/c, their discovery may be favored,
therefore 8\% should be regarded as an upper limit for their frequency
of occurence. For a ``Milky-Way--like'' galaxy (i.e. $L_B=2.3\times
10^{10}L_{B_\odot}$; $M=9.5\times 10^{10}M_\odot$; and morphological
Hubble type Sbc, data from Cox 2000) we obtain a rate of type Ib/c SNe
of $\sim 3.2\times 10^{-3}$ yr$^{-1}$ (after assuming a rate of 0.14
SNe per century and per 10$^{10}$ $L_{B,\odot}$; Cappellaro, Evans \&
Turatto 1999), and therefore the Hypernova rate turns out to be $\sim
2.6\times 10^{-4}$ yr$^{-1}$. This rate has to be compared with the
rate of GRBs in the Milky Way. This quantity can be estimated by
combining the local rate of 0.5 GRB event Gpc$^{-3}$ yr$^{-1}$
(Schmidt 2001), the local density of B luminosity of $\sim 1.2\times
10^8 L_{B,\odot}$ per $Mpc^3$ (e.g. Madau, Della Valle
\& Panagia 1998) and the B luminosity of the Milky Way ($2.3\times
10^{10}L_{B_\odot}$). This approach gives $R_{GRB}\sim 3.8\times
10^{-7}$ yr$^{-1}$ that has to be rescaled for the beaming factor
$f_b^{-1}$. There exist different estimates for this parameter: from
$\sim 500$ (Frail et al. 2001) to $\sim 75$ (Guetta, Piran \& Waxman
2005, Piran 2005). These figures implies that the ratio GRB/Hypernovae
spans the range $\sim 0.7\div 0.11$. Following Mannucci et al. (2005),
who provide the SN rates normalized to the mass in stars of the host
galaxies, the same kind of computation yields a ratio GRB/Hypernovae
of $\sim 0.20\div 0.03$.  These data as a whole do not support a ratio
GRB/Hypernova=1, unless to assume large values of $f_b^{-1}$
(e.g. Frail et al. 2001, Yonetoku et al. 2005). A piece of evidence in
this direction comes from observations of the radio properties
exhibited by SN 2002ap (Berger, Kulkarni \& Chevalier 2002) that do
not support the association of this Hypernova with a GRB.  For
$f_b^{-1}\sim 50-100$ (Guetta et al. 2004, Piran 2005) the ratio
GRB/Hypernova should be of the order of $\sim 0.1$ (or even less).
Incidentally we note that the ratio GRBs/SNe-Ibc $\sim 0.05\div 0.008$
(which can be obtained from the rates reported above) is consistent
with the results independently obtained by Berger et al. 2003 (see
also Soderberg 2005), who derived, from radio observations of 33
`local' SNe-Ib/c, an incidence of 1998bw-like events over the total
number of SNe-Ibc of $<0.03$ (see also Granot \& Ramirez-Ruiz 2004).

\begin{table}
  \caption{Supernova-Gamma Ray Burst time lag. A negative time lag indicates
  that the SN explosion precedes the GRB.}  
\label{}
 \begin{narrowtabular}{2cm}{rrrcr} 
GRB & SN & $+\Delta {\rm t(days)}$ &  $-\Delta {\rm t(days)}$ & References \\ 
\hline 
GRB 980425 & 1998bw & 0.7 &  --2 & Iwamoto et al. 1998\\ 
GRB 000911 & bump & 1.5 & --7 & Lazzati et al. 2001\\ 
GRB 011121 & 2001ke & 0 & --5 & Bloom et al. 2002b \\
           &      & --  & a few&Garnavich et al. 2003\\  
GRB 021211 & 2002lt & 1.5 & --3 & Della Valle et al. 2003\\ 
GRB 030329 & 2003dh & 2 &  --8  & Kawabata et al. 2003\\ 
           &        & --  &  --2  &Matheson et al. 2003\\ 
GRB 031203 & 2003lw & 0 & --2 & Malesani et  al. 2004\\ 
\hline \end{narrowtabular}
\end{table}

\section{SN-GRB time lag}

Several authors have reported the detection of Fe and other metal
lines in GRB X-ray afterglows (e.g. Piro et al. 1999; Antonelli et
al. 2000; Reeves et al. 2002). If valid (see Sako et al. 2005 for a
critical view) these observations would have broad implications for
both GRB emission models and would strongly link GRBs with SN
explosions. For example, Butler et al. (2003) have reported the
detection in a Chandra spectrum of emission lines whose intensity and
blueshift would imply that a supernova occurred $>2$ months prior to
the $\gamma$ event. This kind of observations can be accommodated in
the framework of the {\sl supranova} model (Vietri \& Stella 1998),
where a SN is predicted to explode months or years before the $\gamma$
burst. In Tab. III we have reported the estimates of the lags between
the SN explosions and the associated GRBs, as measured by the authors
of the papers.  After taking the data of Tab. III at their face value,
it is apparent that the SNe and the associated GRBs occur
simultaneously, providing support to the {\sl collapsar} scenario
(Woosley 1993, Paczyski 1998, Mac Fadyen \& Woosley 1999).  Data in
Tab. III provide useful constraints to those models that predict the
SNe to occur before (Vietri \& Stella 1998) or after (Ruffini et
al. 2001) the gamma-ray event.

\section{\ldots There is an Expanding Frontier of Ignorance \ldots
{\scriptsize {(R. Feynman, {\sl Six Easy Pieces})}}}

Data presented in previous sections provide robust empirical grounds to the
idea that some types of core-collapse SNe are the progenitors of
long-duration GRBs (see also Dar 2004, Matheson 2005, Stanek 2005). On
the other hand, the existence of SN/GRB associations poses intriguing
questions which have not yet been answered.
\smallskip

{\bf 1.} {\sl What kind of SNe are connected with long-duration GRBs
and XRFs?}  
\smallskip

Evidence based on the associations between SN\,1998bw/GRB\,980425,
SN\,2003dh/ GRB\,030329, and SN\,2003lw/GRB\,031203 would suggest that
the parent SN population of GRBs is formed by the bright tail of
Hypernovae. However, there is growing evidence that other types of
SNe-Ib/c, such as `standard' {\sl Ic} events like SN 1994I or faint
Hypernovae can contribute to produce GRBs/XRFs (Della Valle et
al. 2003; Fynbo et al. 2004, Fig. 7; Tominaga et al. 2004;
Levan et al. 2005a, Fig. 8; Masetti et al. 2003, Price et al. 2003a,
Price et al. 2003b, Soderberg et al. 2005, Gorosabel et
al. 2005). Available data suggest the existence of $\sim 5$ mag spread
($M_V\sim -16\div -21$) in the absolute magnitudes at maximum of
SNe-Ib/c associated with GRBs/XRFs, which may be similar to the
magnitude spread exhibited by local SNe-Ib/c (see Fig. 11 in Soderberg
et al. 2005).

Possible associations between GRBs and other types of core-collapse
SNe (particularly with type IIn) have been claimed in the past on
the basis of spatial and temporal SN/GRB coincidences by Germany et
al. (2000) and Turatto et al. (2000) (SN 1997cy/GRB 970514) and by
Rigon et al. (2003) (SN 1999E/GRB 980910). However, in a recent study,
Valenti et al. (2005) were not able to confirm these associations to
be statistically significant (see also Wang \& Wheeler 1998; Kippen et
al. 1998). Currently the best evidence for the case of an association
between a Supernova IIn and a gamma ray burst has been provided by
Garnavich et al. (2003) who find that the color evolution of the bump
associated with GRB 011121 is consistent with the color evolution of
an underlying SN (dubbed as SN 2001ke) strongly interacting with a
dense circumstellar gas due to the progenitor wind (as confirmed 
by radio observations, see Price et al. 2002).
\smallskip

\begin {figure}[!h]
\centering
\begin{center}
\includegraphics [width=7cm, height=7.0cm,angle=0]{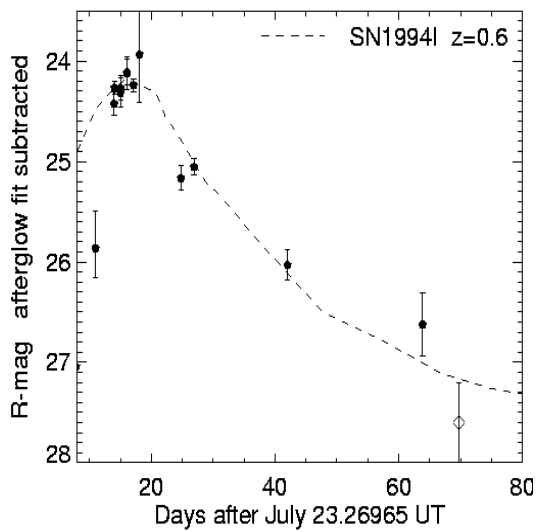}
\caption{The light curve of the bump associated with XRF 030723
compared with the B light curve of the `standard' (type-Ic) SN 1994I
redshifted to $z=0.6$ (data from Richmond et al. 1996). Plot from
Fynbo et al. 2004.}
\label{figura}
\end{center}
\end{figure}

\begin {figure}[!h]
\centering
\begin{center}
\includegraphics [width=7.9cm, height=8.5cm,angle=270]{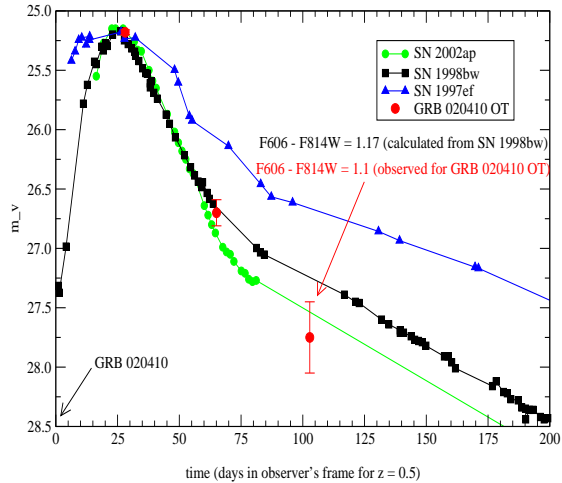}
\caption{Light curve of the SN associated with GRB 020410
compared with SN 1997ef (Iwamoto et al. 2000), 2002ap (Gal-Yam et
al. 2002, Pandey et al. 2003) and SN 1998bw (Galama et al 1998,
McKenzie \& Schaefer 1999). Plot from Levan et al. 2005a.}
\label{figura}
\end{center}
\end{figure}

{\bf 2.} {\sl What are the most frequent gamma-ray events in the
Universe?}  
\smallskip

GRB\,031203 was quite similar to GRB\,980425, albeit more
powerful. Both events consisted in a single, under-energetic
pulse. Their afterglows were very faint or absent in the optical, and
showed a very slow decline in the X rays. Moreover, they were both
accompanied by a powerful Hypernova. Therefore, GRB\,980425 can no
longer be considered as a peculiar, atypical case. Both bursts were so
faint, that they would have been easily missed at cosmological
distances. Since the volume they sample is $10^5 \div 10^6$ times
smaller than that probed by classical distant GRBs, the rate of these
events could be dramatically larger, perhaps they are the most common
GRBs in the Universe.  However we are still left with the question of
whether or not these bursts belong to a different local population of
$\gamma$-bursts (Bloom et al. 1998, Soderberg, Frail \& Wiering 2004)
or they are typical cosmological bursts observed off-axis (Nakamura
1999; Eichler \& Levinson 1999; Woosley et al. 1999; Ramirez-Ruiz et
al. 2004; see also Waxman 2004 for a {\sl pro {\rm and} con}
discussion). Naively one may expect that the spectroscopic and
photometric similarities exhibited by SNe 1998bw, 2003dh and 2003lw
may indicate a common origin for the associated GRBs, in spite of the
fact that they have exhibited dramatic differences in their
$\gamma$-energy budgets and in the properties of their afterglows. We
note that this inference is supported by statistical arguments
provided by Guetta et al. (2004).
\smallskip

{\bf 3.}{\sl What is the relationship between the SN magnitudes at
maximum light and the gamma-ray energy budget?}
\smallskip

Simple statistical analysis of data points reported in Fig. 9 shows
that the absolute magnitude at maximum of SNe associated with GRBs
does not appear to correlate with the respective gamma energy
(although this conclusion should be taken with caution because of the
usage of scanty statistic). The distribution of the data points in
Fig. 9 also reflects an obvious bias, namely the detection of
over-bright SNe is favored because most GRBs are discovered at
cosmological distances and/or their bright afterglows can easily
outshine `standard' SNe-Ib/c or faint Hypernovae. However it is not
clear if the lack of faint SNe associated with intrinsically faint
(and nearby) GRBs is the result of an exiguous statistic or this
finding has a deeper physical meaning.
\smallskip

\begin {figure}[!h]
\centering
\begin{center}
\includegraphics [width=8.4cm, height=8.4cm,angle=0]{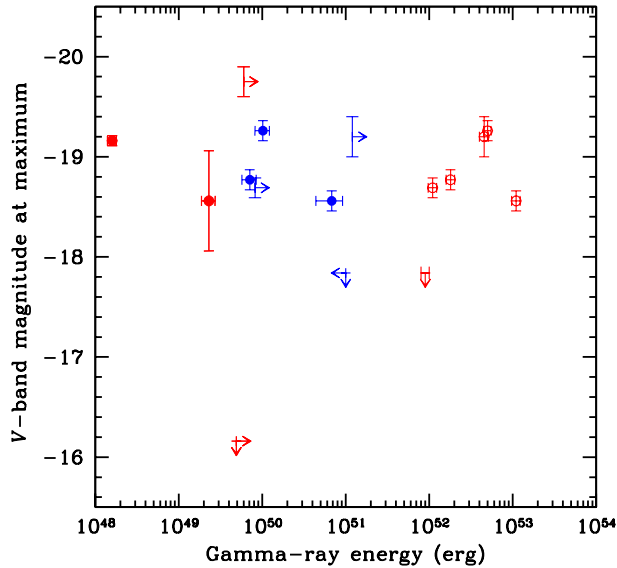}
\caption{
Absolute magnitude of the SN vs. energy gamma budget, assuming
isotropic emission (light red points) and corrected for beaming (blue
points). Red symbols below $10^{50}$ erg (i.e. GRB 980425/SN1998bw;
GRB 031203/SN 2003lw; XRF 020903 and XRF 040701 have not been
corrected for beaming). Data for 1998bw/980425 (Galama et al. 1998);
2002lt/021211 (Della Valle et al. 2003); 2003dh/030329 (Matheson et
al. 2003a); 2003lw/031203 (Malesani et al. 2004); 2001ke/011121
(Garnavich et al. 2003); bump/020405 (Masetti et al. 2003); GRB 010921
(Price et al. 2003b); XRF 020903 and XRF 040701 (Soderberg et
al. 2005); GRB 041006 (Stanek et al. 2005). Isotropic and corrected
for beaming energies are taken from Ghirlanda, Ghisellini \& Lazzati (2004, and
references therein).}
\label{figura}
\end{center}
\end{figure}

{\bf 4.} {\sl May it be that GRBs, which occur in the inner and
outer regions of hosts, have different progenitors?}
\smallskip

Ramirez-Ruiz, Lazzati \& Blain (2002) found some evidence that outer
bursts appear to have systematically greater isotropic equivalent
energies (or narrower jets). These results may be interpreted in terms
of different environmental properties, between inner and outer regions
of the hosts (e.g. metallicity, fraction of binary systems), which can
affect the evolution of the progenitors of core-collapse SNe (see
Bressan, Della Valle \& Marziani 2002 for a discussion).
\smallskip

{\bf 5.} {\sl Are the ``red bumps'' always representative of the
signatures of incipient SNe?} 
\smallskip

Or can some of them be produced by different phenomena such as dust
echoes (Esin \& Blandford 2000) or thermal re-emission of the
afterglow light (Waxman \& Draine 2000). To date, only for
GRB\,021211/SN 2002lt (Della Valle et al. 2003) and XRF 020903
(Soderberg et al. 2005) a spectroscopic confirmation was obtained. On
the other hand, Garnavich et al. (2003) and Fynbo et al. (2004) did not
find SN (spectroscopic) features in the bumps of GRB\,011121 and
XRF\,030723 (see Butler et al. 2005 for an alternative interpretation
of the bump discovered in XRF\,030723).
\smallskip

{\bf 6.} {\sl Is the lack of an optical bump indicative of the lack of
a supernova?} 
\smallskip

Price et al. (2003b), Levan et al. (2005b), Soderberg et al. (2005)
have carried out unsuccessful HST searches for SN signatures in
GRBs/XRFs light curves. For example Soderberg et al. (2005) were able
to set a firm upper limit to the magnitude of the SN associated with
XRF 040701, of $M_V\lsim -16.2$ (see the faintest upper limit in
Fig. 9).  This behaviour can be explained in a number of ways: {\sl
i)} the SN parent population of GRBs/XRFs is formed by a heterogeneous
class of objects that span (at maximum light) a broad range of
luminosities (about a factor 100); {\sl ii)} some `long'' GRBs may
originate by merging compact objects (e.g. Belczynski, Bulik \& Rudak
2002; Della Valle, Marziani \& Panagia 2005) rather than in SN
explosions; {\sl iii)} sometimes SN and GRB do not occur
simultaneously (e.g. Vietri \& Stella 1998). For a delay of a few
weeks/months, the supernova would have faded before the GRB was
detected, and this may explain why supernovae are not discovered after
every GRB. Finally we note that the light curve of the afterglow of
GRB 030329 (associated with SN 2003dh) did not show the bump which is
believed to be caused by the emerging SN (Matheson et al. 2003a their
Fig. 13, Lipkin et al. 2004).
\smallskip
                                                               
{\bf 7.} {\sl What causes some small fraction of SNe Ib/c to produce
observable GRBs, while the majority do not?}  
\smallskip

With the obvious exceptions of SN 1998bw, 2003dh and 2003lw, none of
the Hypernovae reported in Tab. II have been associated with GRBs by
direct observations (the association GRB 971115/SN 1997ef, for
example, has been proposed by Wang \& Wheeler (1998) on the basis of
spatial and temporal coincidences).  The situation is even more
intriguing if one considers that Hypernovae are only a small fraction
of `normal' SNe-Ibc (less than 10\%) and thus only a very tiny
fraction of SNe-Ib/c, about $0.8\% \div 5\%$, seems to be able to
produce GRBs. This may imply that the evolution leading a star to
produce a GRB requires very special circumstances (e.g. rotation,
binary interaction; see Podsiadlowski et al. 2004, Mirabel 2004, Fryer
\& Heger 2005) other than being `only' a very massive star. As an
alternative, one can argue that most SNe Ib/c (if not all of them)
produce GRBs (Lamb, Donaghy \& Graziani 2005). This fact would imply
very small jet opening angles ($\sim 1$ degree) and therefore one
should be able to detect as GRBs only those events which are viewed at
very small angles relative to the jet direction. More spherically
symmetric jets or events viewed from angles (relative to the jet
direction) which are larger than the typical viewing angles of
long-duration GRBs, should yield XRFs (see also Dado, Dar \& De Rjula
2004). Finally, at even larger angles, relative to the jet direction,
one should observe `only' the SN(Ib/c) explosions. With a rate of
discovery of about 2 event/week, the {\sl Swift} satellite (Gehrels et
al. 2004) will allow the GRB community to obtain in the next $3 \div
4$ years an accurate spectroscopic classification for dozens SNe
associated with GRBs and to provide conclusive answers to several of
the above questions.
\acknowledgments

It is a pleasure to thank N. Panagia and D. Malesani for their useful
comments and the critical reading of the manuscript. I'm also indebted
with S. Benetti, G. Chincarini, F. Frontera, N. Gehrels, K. Hurley,
P. Mazzali, L. Stella, G. Tagliaferri and V. Trimble for suggestions
and illuminating discussions and with M. Hamuy, K. Stanek, A. Levan,
J. Fynbo and F. Patat to have made available their plots.

\end{document}